\documentclass[12pt]{iopart}

\usepackage{iopams}  
\usepackage{graphicx}
\begin{document}
\title{Geometric analysis of entangled qubit pairs}

\author{
A. Ram\v{s}ak}
\address{Faculty of Mathematics and Physics, University of
Ljubljana, Ljubljana, Slovenia}
\address{J. Stefan Institute, Ljubljana, Slovenia}

\date{\today}

\begin{abstract}
Two entangled electron spins, or qubits, are analyzed in terms of ordinary three-dimensional space geometric properties, as are the angles between their angular momenta. This formulation allows concurrence, a measure of quantum entanglement, to be expressed as expectation values of trigonometric functions of the azimuthal angle between the two angular momenta.
\end{abstract}

\submitto{\NJP}

\maketitle

As quantum entanglement does not have a direct classical analogue and as it is not an observable according to the usual rules of quantum mechanics, it is not obvious how it might be visualized beyond the standard formalism. Nevertheless, several approaches have
so far been developed in order to intuitively analyse this problem in terms of some geometrical quantities. For example, the entanglement of two spin-1/2 particles can be studied in terms of the distances between states in high-dimensional manifolds \cite{kus}, one can
define entanglement geometrically as the distance between a given state and the nearest separable state \cite{wei03,uyanik}, relate
entanglement to the geometrical structure of the quaternionic Hopf fibration \cite{levay}, or analyse it by the approach of the operator trigonometry  \cite{gustafson}. A review of separability criteria and entanglement measures is discussed geometrically in Ref.~\cite{krenaknjiga}. These geometric approaches rely predominantly on abstract higher dimensional spaces not on usual $\Bbb{R}^3$. 

Indeed, in general, quantities relating to spin-1/2 systems cannot be described in terms of classical variables, in particular not in terms of usual geometrical coordinates. Still, to some extent,  geometric imagery is possible for a spin-1/2 state of a single electron which is customarily
parametrized in terms of the Bloch or Poincar\' e sphere, where polar and azimuthal angles can be interpreted as the Euler angles of a unity vector pointing along the spin direction. One possible generalisation of the Bloch sphere is the sphere model, which gives a geometrical view of entanglement in terms of constraint functions describing the behaviour of the state of one of spins if measurements are made on the other \cite{aerts}. 

The aim of the present paper is firstly to visualize  the quantum state of a pair of spin-1/2 particles geometrically in ordinary three-dimensional space and, secondly, to express the quantum entanglement measure concurrence \cite{wootters} intuitively in terms of angles made by their angular momenta.

The azimuthal angle $\phi$ and the orbital angular momentum $L_z=-i \partial /\partial \phi$ are commonly regarded as conjugate variables connected by the commutator $[\phi,L_z]=i$. However, the use of this angle variable requires avareness because  the position in space is a periodic  function of $\phi$, which itself is not periodic \cite{carruthers63}. A similar problem appears in the treatment of the phase variable in a quantum mechanical harmonic oscillator. In particular, it was shown that Dirac's \cite{dirac27} assumption of a Hermitian operator $\hat\phi$ conjugate to the number operator leads to contradictions \cite{susskind64}. 
A proper description of an angle or a phase variable requires that periodicity be taken into ccount. It turns out that a simple way of locating the azimuthal position is to give $\cos \phi$ and $\sin \phi$ instead of $\phi$ \cite{louisell63}. These operators are related by commutators $[\sin \phi,L_z]=i\cos\phi$ and $[\cos \phi,L_z]=-i\sin\phi$ \cite{carruthers63}.

Here we investigate the role of azimuthal coordinates of two entangled electrons, or qubits. We apply this picture to the case of a qubit pair described by a mixed state and then quantify the quantum entanglement of the pair \cite{trije}. 
Qubits are not restricted to real spin of electrons  and can represent any two state quantum system, for example,  entangled photons pairs  \cite{fotoni}, flux qubits in superconducting rings  \cite{makhlin},
charge qubits in double quantum dots  \cite{mravlje},  flying qubits in quantum point contacts \cite{rejec},  or two-qubit systems with different types of particles  \cite{delfot}.

First we consider two qubits in a state with a vanishing total spin projection,
\begin{equation}
| \Psi\rangle=\cos { \vartheta \over 2} \left|\uparrow\downarrow\right\rangle+e^{i \varphi}\sin {\vartheta \over 2} \left|\downarrow\uparrow\right\rangle.
\label{psi2}
\end{equation}
For convenience we use  the notation
where $\left|\uparrow\downarrow\right\rangle$, for example, represents the state where the first particle (qubit) is in the "up" state, {\it i.e.}, in the direction of the $z$-axis, and the second qubit is in the state "down". For the Schmidt angle $\vartheta/2$ we assume $\vartheta\in[0,\pi]$.
Spin operators are given by
$\mathbf{S}_{1(2)}={1\over2}(\sigma_{1(2)x},\sigma_{1(2)y},\sigma_{1(2)z})$ for the first and the second qubit, respectively, and $\sigma$ are the Pauli matrices. 

\begin{figure}[htbp]
\begin{center}
\includegraphics[width=55mm]{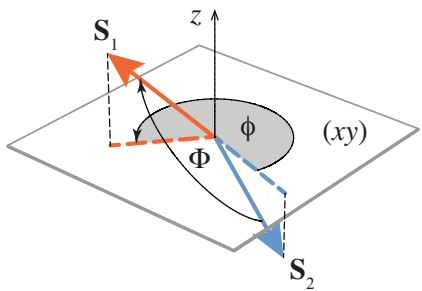}
\caption{Visualization of the angle $\Phi$ between the angular momenta of the first and the second qubit and the difference of azimuthal angles, 
made by the angular momenta projections onto the $xy$-plane, $\phi=\phi_1-\phi_2$.}
\label{Fig1}
\end{center}
\end{figure}

Geometric view in ordinary three-dimensional space of such an entangled qubit pair is gained by the analysis of relative angles
between the angular momenta. There are two main alternatives. The first is the angle $\Phi$ made of the angular momenta as shown in Fig.~\ref{Fig1}. As the expectation value of the corresponding cosine we take
\begin{equation}
\left\langle \cos\Phi \right\rangle=\langle\Psi| {\mathbf{S}_{1}\cdot  \mathbf{S}_{2} \over \sqrt{\mathbf{S}_{1}^2\mathbf{S}_{2}^2}}|\Psi \rangle=
{1 \over 3}(2\sin\vartheta\cos\varphi-1),
\label{ffi}
\end{equation}
with the variance $\Delta \cos\Phi=\frac{2}{3}\sqrt{1-\sin^2\vartheta\cos^2\varphi}$ vanishing for the singlet or the triplet state.
This is a well known result, where the angular momenta can be visualized as antiparallel for the case of the singlet and not quite parallel  for the triplet qubit pair state where $\Phi\approx 71^\circ$.

In this paper we concentrate on the properties of another geometric element, the azimuthal angle difference $\phi=\phi_1-\phi_2$ made of angular momenta projections onto the $xy$-plane, Fig.~\ref{Fig1}. Here $\phi_{1(2)}$ can be visualized as the  azimuthal angle of the first or the second qubit, respectively. We define two {\it operators} corresponding to $\cos\phi$ and $\sin\phi$,
\begin{eqnarray}
\widehat{\cos\phi}&\equiv&{{S_{1x} S_{2x}+S_{1y} S_{2y}} \over  
\sqrt{(S_{1x}^2 +S_{1y}^2)(S_{2x}^2 +S_{2y}^2)} }, \\ 
\widehat{\sin\phi}&\equiv&{{S_{1y} S_{2x}-S_{1x} S_{2y}} \over  
\sqrt{(S_{1x}^2 +S_{1y}^2)(S_{2x}^2 +S_{2y}^2)} }.
\label{cossin}
\end{eqnarray}
These operators obey commutator relations analogous to the orbital momentum case  \cite{louisell63,carruthers63},
\begin{eqnarray}
\lbrack{\widehat{\sin\phi},\delta S_z}\rbrack&=&i \widehat{\cos\phi}, \label{comm1}\\
\lbrack\widehat{\cos\phi},\delta S_z\rbrack&=&-i \widehat{\sin\phi}, 
\label{comm2}
\end{eqnarray}
where the relative angular momentum $\delta S_z=S_{1z}-S_{2z}$ plays the role of $L_z$ in the case of orbital motion, which supports the view that $\phi$ corresponds to the relative azimuthal angle.

There are two pairs of eigenstates and eigenvalues for these operators,
\begin{eqnarray}
\widehat{\cos\phi}\;( \left|\uparrow\downarrow\right\rangle\pm\left|\downarrow\uparrow\right\rangle)&=&\pm (\left|\uparrow\downarrow\right\rangle\pm\left|\downarrow\uparrow\right\rangle), \label{eigen1}\\ 
\widehat{\sin\phi}\;(\left|\uparrow\downarrow\right\rangle\pm i \left|\downarrow\uparrow\right\rangle)&=&\pm (\left|\uparrow\downarrow\right\rangle\pm i \left|\downarrow\uparrow\right\rangle).
\label{eigen2}
\end{eqnarray}
For the triplet and singlet state the projections of angular momenta make angles $0$ and $\pi$, respectively, and the momenta for eigenstates of 
$\widehat{\sin\phi}$ are perpendicular. 

Note that these results imply that there are only four possible values for angle $\phi$, contrary to the naive intuition that $\varphi$ in equation~(\ref{psi2}) corresponds to $\phi$, at least for the perfectly entangled qubit pairs, i.e., states with $\vartheta=\frac{\pi}{2}$. Such an interpretation follows from a heuristic argument: consider a triplet state $\left|\uparrow\downarrow\right\rangle+\left|\downarrow\uparrow\right\rangle$, where  angle $\phi$ is zero. Then by the unitary transformation $\exp(-i \varphi S_{1z})$ rotate the first spin around the $z$-axis  for angle $\varphi$. The outcome is the state proportional to $|\Psi\rangle$, therefore one would expect that the angular momenta then make angle $\phi=\varphi$. Below will be given alternative arguments that this view indeed can be considered correct.

As discussed for the case of orbital motion, a proper definition of angle  operator $\hat\phi$ is  impossible due to the problem of periodicity \cite{dirac27,carruthers63,note1}. However, one can define a pair of operators as in ordinary trigonometry,
\begin{eqnarray}
\hat \phi_c&\equiv&\arccos \widehat{\cos\phi}=\frac{\pi}{2}(1-\widehat{\cos\phi}), \\ 
\hat \phi_s&\equiv&\arcsin\widehat{\sin\phi}=\frac{\pi}{2}\widehat{\sin\phi}.
\label{koti}
\end{eqnarray}
The operator $\hat \phi_c$ is actually the operator for the "absolute value of the angle", $\hat \phi_c\equiv\widehat{|\phi |}$, due to the periodicity properties of trigonometric functions defined on the interval $[0,\pi]$. $\hat \phi_c$ does not represent any new information as it can be by the Taylor expansion and the known properties of the Pauli matrices reexpressed in terms of $\widehat{\cos\phi}$. The operator $\hat \phi_s$ is expressed by $\widehat{\sin\phi}$ and also can
not be considered as a proper angle operator, because the inverse sine maps onto $[-\frac{\pi}{2},\frac{\pi}{2}]$. In total there are four possible eigenvalues for the angle operators: $\phi=0$, $\pi$ and $\pm\frac{\pi}{2}$ for $\hat \phi_c$ and $\hat \phi_s$, respectively, as expected for the four eigenvectors in equations~(\ref{eigen1}) and (\ref{eigen2}).

The interpretation of the expectation values for the angle operators would require, due to the periodicity property, the resolving of a double mapping, which can not be done in a unique way. It is sufficient to study the cosine and sine operators and a direct evaluation reveals
\begin{eqnarray}
\langle\Psi|\widehat{\cos\phi}|\Psi \rangle& =&C \cos \varphi, \label{c}\\ 
\langle\Psi|\widehat{\sin\phi}|\Psi \rangle &=&C \sin \varphi,
\end{eqnarray}
where $C=\sin\vartheta$. For perfectly entangled qubit pairs these two equations simplify to intuit forms, $\left\langle\widehat{\cos\phi}\right\rangle=\cos \varphi$  and $\left\langle\widehat{\sin\phi}\right\rangle=\sin \varphi$, which in this formulation supports the anticipated view that $\phi=\varphi$ for fully entangled states.  Identical results were found also in a recent analysis of quantum entanglement in the de Broglie-Bohm interpretation of quantum mechanics \cite{geometrical}.

Quantum entanglement can be quantified by  various measures \cite{1995}, one of them being
the entanglement of formation \cite{bennett} which is the asymptotic conversion rate to maximally entangled states from an ensemble of  copies of a non-maximally entangled state  \cite{vedral}. By the Wootters formula \cite{wootters} is the entanglement of formation simply related to an associated quantity, the concurrence $C$, introduced in equation~(\ref{c}).

Entanglement of two qubits can therefore be visualized in terms of functions of a simple geometric quantity -- the difference of azimuthal angles of the angular momenta -- by a  simple relation 
\begin{equation}
C=\sqrt{
\left\langle\widehat{\cos\phi} \right\rangle
^2 +\left\langle\widehat{\sin\phi }
\right\rangle^2 }.
\label{cc}
\end{equation}
This expression offers a clear interpretation that the concurrence is small if the angles are spread randomly, leading to small cosine and sine averages. The concurrence is close to unity if angles are packed at some common angle difference $\varphi$ such that the sum of squared averages of cosine and sine adds close to unity. As a measure of angle spreading can serve the variances for cosine and sine operators, to $C$ related by the expression
 \begin{equation}
(\Delta \widehat{\cos\phi})^2+(\Delta \widehat{\sin\phi})^2=2-C^2.
\label{dde}
\end{equation}
Larger concurrence thus signals lower fluctuations of cosine and sine. However, even for perfectly entangled states with $C=1$ the sum of variances (\ref{dde}) is non-zero, because $|\Psi\rangle$ is never an eigenstate of cosine and sine operators simultaneously \cite{cossin3}.

The state $|\Psi\rangle$ is related to the state
\begin{equation}
|\widetilde{ \Psi}\rangle=\cos { \tilde\vartheta \over 2}\left |\uparrow\uparrow\right\rangle+e^{i \tilde\varphi}\sin {\tilde\vartheta \over 2} \left|\downarrow\downarrow\right\rangle
\label{tpsi2}
\end{equation}
where the spin of the second qubit is reversed by the unitary transformation $\exp(-i \pi S_{2y})$,  and $\vartheta \to \tilde\vartheta$, $\varphi+\pi \to \tilde\varphi$. The formalism based on cosine and sine operators can be applied in a similar manner as for the case of $\left |\Psi\right\rangle$. The main distinction is, that for states (\ref{tpsi2})  the sum of azimuthal angles, $\tilde \phi=\phi_1+\phi_2$ is important, not the difference $\phi$. The corresponding trigonometric operators
\begin{eqnarray}
\widehat{\cos\tilde\phi}&\equiv&{{S_{1x} S_{2x}-S_{1y} S_{2y}} \over  
\sqrt{(S_{1x}^2 +S_{1y}^2)(S_{2x}^2 +S_{2y}^2)} }, \\ 
\widehat{\sin\tilde\phi}&\equiv&{{S_{1y} S_{2x}+S_{1x} S_{2y}} \over  
\sqrt{(S_{1x}^2 +S_{1y}^2)(S_{2x}^2 +S_{2y}^2)} }
\label{cossin2}
\end{eqnarray}
together with $S_z=S_{1z}+S_{2z}$ exhibit similar commutation relations to Eqs.~(\ref{comm1}), (\ref{comm2}). All  results valid for $\phi$ can be mutatis mutandis applied to $\tilde\phi$ and are not shown here. We present here only the concurrence $\widetilde C=\sin\tilde\vartheta$,
\begin{equation}
\widetilde{C}=\sqrt{
\left\langle\widehat{\cos\tilde \phi}\right \rangle
^2 +\left\langle\widehat{\sin\tilde \phi }
\right\rangle^2 }.
\label{tcc}
\end{equation}
It should be noted that $C=0$ for $|\tilde\Psi\rangle$ and $\widetilde C=0$ for $|\Psi\rangle$.

Finally, we can express the concurrence $C_\rho$ for a system of two qubits in a mixed state given by the density matrix 
\begin{equation}
\rho=\sum_i p_i |\Psi_i\rangle\langle\Psi_i|+\sum_j\tilde p_j |\widetilde\Psi_j\rangle\langle\widetilde\Psi_j|,
\end{equation}
which represents systems with a conserved square of total spin projection, since $[\rho,S_z^2]=0$. For this class of systems the concurrence is given by \cite{amico08,ramsak1}
\begin{equation}
C_\rho = 
\max\left(0,C-2\sqrt{\langle
P_{1}^{\uparrow}P_{2}^{\uparrow}\rangle\langle
P_{1}^{\downarrow}P_{2}^{\downarrow}\rangle}, \widetilde C-2\sqrt{\langle
P_{1}^{\uparrow}P_{2}^{\downarrow}\rangle\langle
P_{1}^{\downarrow}P_{2}^{\uparrow}\rangle}\right),\label{ccc}
\end{equation}
where $P^m_{1(2)}=\frac{1}{2}\pm S_{1(2)z}$ is  the projector
onto the spin state $m=\uparrow, \downarrow$, respectively.  The expectation values are evaluated as is customary for mixed states, for example, $\left\langle \widehat{\cos\phi}\right\rangle=\mathrm{Tr} \rho \widehat{\cos\phi}$ or $\langle P_{1}^{\downarrow}P_{2}^{\downarrow}\rangle=\sum_j  \tilde p_j \sin^2 \frac{\tilde\vartheta_j}{2}$.

To summarize, we have analyzed a pair of entangled qubits  in terms of observables expressed by trigonometric functions of relative angles made of spin vectors of two qubits. Following the approach developed for the analysis of angles and angular momenta of orbital motion we applied analogous operators corresponding to cosine and sine of azimuthal angles difference $\phi=\phi_1-\phi_2$ for the case of spin-1/2. In terms of such operators we constructed appropriate azimuthal angle operators, which reveals that for states with vanishing total spin projection there are four eigenstates of angle operators, with eigenvalues $\phi=0, \pm\frac{\pi}{2}, \pi$. The analysis of the expectation values of trigonometric operators gives support to an argument that the phase factor $\varphi$ in the wave function (\ref{psi2}) corresponds to $\phi$. 

Average cosine and sine of this angle are plainly related to the concurrence, a measure of the degree of quantum entanglement. 
In particular, the concurrence is exactly given by  $C=\left\langle\widehat{\cos\phi}\right\rangle$, for $\varphi=0$, and is also related to the corresponding variance $\Delta \widehat{\cos\phi}=\sqrt{1-C^2}$. A higher degree of entanglement can thus be visualized as a highly correlated distribution of angular momenta making azimuthal angles close to $\varphi$ and with suppressed fluctuations with progressively increasing entanglement.
An analogous analysis was performed for the space spanned by the basis vectors $\left|\uparrow\uparrow\right\rangle$ and $\left|\downarrow\downarrow\right\rangle$. Here the relevant geometric quantity is the sum of azimuthal angles, $\phi_1+\phi_2$, and the concurrence is given in terms of the average cosine of this angle.

The analysis of qubits in pure states was generalized to systems described by mixed states and as the final result the concurrence is expressed in terms of trigonometric operators for a rather general class of systems. Although in any explicit quantification of the entanglement the expressions reduce to the  manipulation of ordinary Pauli spin operators,
we believe that the present approach of the application of geometrical quantities offers a new insight into the phenomenon of quantum entanglement, which is due to the lack of  a direct classical analogue not recognized in standard formal approaches.

The author thanks T. Rejec, J. H. Jefferson, I. Sega,  and T. Huljev {\v C}ade{\v z} for discussions and  he  acknowledges the support from the Slovenian Research Agency under Contracts No. J1-0747 and P1-0044.


\begin{thebibliography}{99}
\bibitem{trije}{Here we do not introduce any new entanglement measures -- the entanglement is analysed by the Wootters concurrence. The generalisation to multiple qubit states is not considered. Such a generalisation  would in principle be possible but the results would probably be cumbersome and would not lead to a clear visualisation of the entanglement which is the main aim of this paper.}
\bibitem{note1}{Note that the commutator equation $\lbrack{\hat\phi,S_{1z}-S_{2z}}\rbrack=iI$, where $I$ is the identity operator, can not be satisfied by any $4\times4$ matrix $\hat\phi$, contrary to the orbital motion case where $\lbrack{\phi,-i \partial /\partial \phi}\rbrack=i$.}
\bibitem{cossin3}{Note the exact operator relation $\widehat{\cos\phi}^2=\widehat{\sin\phi}^2=I$, where $I$ is the identity operator, leading to $\left\langle\widehat{\cos\phi}^2+\widehat{\sin\phi}^2\right\rangle=2$.}
\end{thebibliography}
\end{document}